\documentclass[fleqn,10pt,twocolumn]{wlscirep}
\usepackage[utf8]{inputenc}
\usepackage[T1]{fontenc}

\usepackage{lineno}

\title{Molecular dynamics with a first-principles-validated universal
machine-learning potential reveals dynamic elementary processes of
growth-related adspecies on GaN(0001)}

\author[1]{Yoshito Takaesu}
\author[2,3,*]{Akira Kusaba}
\author[4]{Junko Ishii}
\author[4]{Shigenori Matsushima}
\author[2]{Yoshihiro Kangawa}
\affil[1]{Interdisciplinary Graduate School of Engineering Sciences,
Kyushu University, Fukuoka 816-8580, Japan}
\affil[2]{Research Institute for Applied Mechanics, Kyushu University,
Fukuoka 816-8580, Japan}
\affil[3]{Institute of Materials and Systems for Sustainability,
Nagoya University, Nagoya 464-8601, Japan}
\affil[4]{National Institute of Technology, Kitakyushu College,
Kitakyushu, Japan}
\affil[*]{kusaba@riam.kyushu-u.ac.jp}


\begin{abstract}
Atomic-scale understanding of the surface elementary processes in
metalorganic vapor phase epitaxy (MOVPE) of GaN has so far relied on
static density-functional-theory (DFT) energetics and on first-principles
molecular dynamics (FPMD) limited to a few tens of picoseconds.
Here we combine FPMD with a universal machine-learning interatomic
potential (MLIP), UMA, to follow the dynamics of growth-related
adspecies on GaN(0001) over time scales inaccessible to purely
first-principles approaches.
FPMD simulations of a GaNH admolecule coexisting with H adatoms reveal
a hitherto unrecognized diffusion mode, in which the N atom lifts the
Ga atom of the GaNH unit off the surface layer during migration, and
show
that the lifted Ga abstracts an H adatom from the surface---events
invisible to static DFT.
Single-point UMA calculations on FPMD snapshots reproduce the
first-principles relative energies along the trajectory
(RMSE $\approx$ 8.5~meV/atom) without any retraining.
Long-time MLIP-based MD (150~ps) then reveals dynamics never observed
within the FPMD window: site-to-site H-adatom hopping, which gates the
migration paths of the growth unit, and reversible dissociation of the
GaNH unit into independently migrating Ga and NH adspecies.
This work constitutes, to our knowledge, the first application of an
MLIP to the molecular dynamics of GaN MOVPE.
\end{abstract}

\begin{document}

\flushbottom
\maketitle
\thispagestyle{empty}
%\linenumbers

% ==========================================================================
\section*{Introduction}
% ==========================================================================

Gallium nitride (GaN) is a key wide-band-gap semiconductor that underpins
solid-state lighting and next-generation power and radio-frequency
electronics \cite{Amano2015,AmanoRoadmap2018}.
Metalorganic vapor phase epitaxy (MOVPE), in which trimethylgallium (TMG)
and ammonia (NH$_3$) are supplied with H$_2$ or N$_2$ carrier gas at
growth temperatures around 1300~K, is the principal technique for
producing device-quality GaN films.
Further improvement of crystalline quality---reduction of point and
extended defects, control of impurity incorporation, and morphology
engineering---increasingly demands an atomic-scale understanding of the
elementary processes occurring on the growing surface: adsorption and
decomposition of precursors, migration of adspecies, and their
incorporation into the crystal lattice.

Considerable theoretical progress has been made in this direction.
Ab initio thermodynamic analyses established that the surface
reconstruction of GaN(0001) under MOVPE conditions depends sensitively on
the growth environment; in particular, a Ga-rich surface decorated with
Ga adatoms is realized under N$_2$ carrier gas, whereas hydrogenated
reconstructions such as 3Ga--H dominate under H$_2$ carrier gas
\cite{Kangawa2001,Kusaba2017JJAP}.
On this basis, static density-functional-theory (DFT) calculations of
reaction pathways clarified how the nitrogen source is incorporated into
the growing crystal: NH$_3$ arriving at the Ga-rich GaN(0001) surface is
decomposed heterogeneously with a surprisingly small activation barrier
($\sim$0.6~eV), and the resulting NH$_x$ units intervene in the
\emph{weak Ga--Ga bonds} formed around Ga adatoms, which act as
self-catalytic reactive sites for N incorporation
\cite{Bui2018JPCC,Bui2019JCG}.
The same concept was extended to vicinal surfaces, where NH units
produced on terraces migrate with a moderate barrier and are trapped at
atomic step edges, providing an atomistic picture of step-flow growth
\cite{Bui2021APSUSC}.
On the realistic mixed surface composed of Ga-adatom and Ga--H motifs
\cite{Kusaba2022APL,Kawka2024JAP},
the diffusion of an isolated NH unit is blocked by hydrogenated regions,
but a \emph{concerted} migration of the NH unit accompanied by a Ga
adatom---an NH--Ga complex moving ``hand in hand''---proceeds with a low
barrier and delivers nitrogen to the step edges \cite{Bui2023APSUSC}.
These studies identified the sequence
NH$_3$ decomposition $\rightarrow$ NH$_x$ intervention in Ga--Ga bonds
$\rightarrow$ transport to step edges $\rightarrow$ Ga--N network
formation as the essential route of GaN growth, with the GaNH complex
(the --Ga--(NH)--Ga-- unit and its mobile form) emerging as an
elementary building block and carrier of the new crystalline layer.

Static total-energy calculations, however, describe the surface at zero
temperature and require reaction pathways to be postulated in advance.
First-principles molecular dynamics (FPMD) has begun to reveal how
qualitatively different the growing surface is at the actual growth
temperature.
Car--Parrinello simulations showed that at 1300~K the Ga adatoms on the
Ga-rich GaN(0001) surface form a two-dimensional liquid-like phase with
strongly enhanced, non-Arrhenius diffusion \cite{Bui2020JJAP}, and
free-energy sampling of NH$_3$ on the same surface clarified
finite-temperature adsorption-site preferences, sequential
dehydrogenation barriers, and the role of released hydrogen, which
hydrogenates the surface and in turn restricts Ga adatom diffusion
\cite{Boero2022APSUSC}.
These findings demonstrate that the MOVPE growth front is an intrinsically
dynamic object whose behavior cannot be inferred from static energetics
alone.

Nevertheless, FPMD is computationally so demanding that accessible time
scales remain limited to a few tens of picoseconds for slab models of
realistic size.
Many of the elementary processes that ultimately govern growth are,
however, expected to occur as rare events on time scales beyond this
window, and the dynamic behavior of the growth front on such time
scales has remained essentially unexplored.
Machine-learning interatomic potentials (MLIPs) offer a route across this
gap: trained on first-principles data, they retain near-DFT accuracy at a
computational cost lower by several orders of magnitude
\cite{Behler2007,Deringer2019,Batzner2022NequIP}.
Most recently, \emph{universal} MLIPs pretrained on massive DFT datasets
spanning broad chemical and structural spaces---such as M3GNet
\cite{Chen2022M3GNet}, MACE-MP \cite{MACE2023}, and UMA \cite{UMA2025}---%
have made it possible to simulate complex systems without constructing a
system-specific training set.
For GaN, MLIP-based molecular dynamics has recently been applied to
crystal growth from the melt using a system-specifically trained
potential \cite{Chen2024CGD}.
Whether universal potentials can quantitatively describe the
chemically intricate environment of a growing GaN surface---metallic
Ga--Ga bonding, polar Ga--N bonds, and N--H/Ga--H species coexisting at
high temperature---has, however, never been examined, and to our
knowledge no MLIP-based molecular dynamics study of GaN MOVPE growth
has been reported.

In this work, we combine FPMD and a universal MLIP to investigate the
dynamics of growth-related adspecies on the GaN(0001) surface,
taking as a model system the GaNH admolecule---the elementary growth
unit and nitrogen carrier identified by the earlier static studies
\cite{Bui2018JPCC,Bui2023APSUSC}---coexisting with H adatoms.
We first perform FPMD simulations with the CASTEP code and observe two
dynamic events that static DFT could not anticipate: the GaNH unit
migrates by a mechanism in which its N atom lifts the Ga atom of the
unit off the surface layer, and the lifted Ga captures a neighboring H
adatom, abstracting it from the surface.
We then validate the universal potential UMA against these
first-principles trajectories by single-point recalculation of FPMD
snapshots, finding that the relative energies along the trajectories are
accurately reproduced.
Finally, exploiting the efficiency of UMA, we extend the simulation to
a time scale unreachable by FPMD (150~ps, one hundred times longer) and
observe elementary processes that never occurred within the FPMD time
window: H adatoms hop between adsorption sites, thereby opening and
closing the migration paths of the growth unit, and the GaNH unit
reversibly dissociates into a Ga adatom and an NH admolecule, which
migrate independently and re-encounter each other to re-form the GaNH
unit.
These results establish the usefulness of universal MLIPs for
GaN growth simulations and provide direct dynamical insight into the
elementary processes underlying MOVPE growth of GaN.

% ==========================================================================
\section*{Methods}
% ==========================================================================

\subsection*{First-principles molecular dynamics}

First-principles molecular dynamics simulations were carried out within
DFT using the plane-wave pseudopotential code CASTEP \cite{CASTEP}.
Exchange and correlation were treated with the PBE generalized gradient
approximation \cite{PBE} without dispersion correction.
Ultrasoft pseudopotentials were employed with a plane-wave cutoff
energy of 280~eV, and the Brillouin zone was sampled at the $\Gamma$
point.
Partial occupancies were treated with Gaussian smearing of width
0.1~eV.

\begin{figure}[!tb]
\centering
\includegraphics[width=\linewidth]{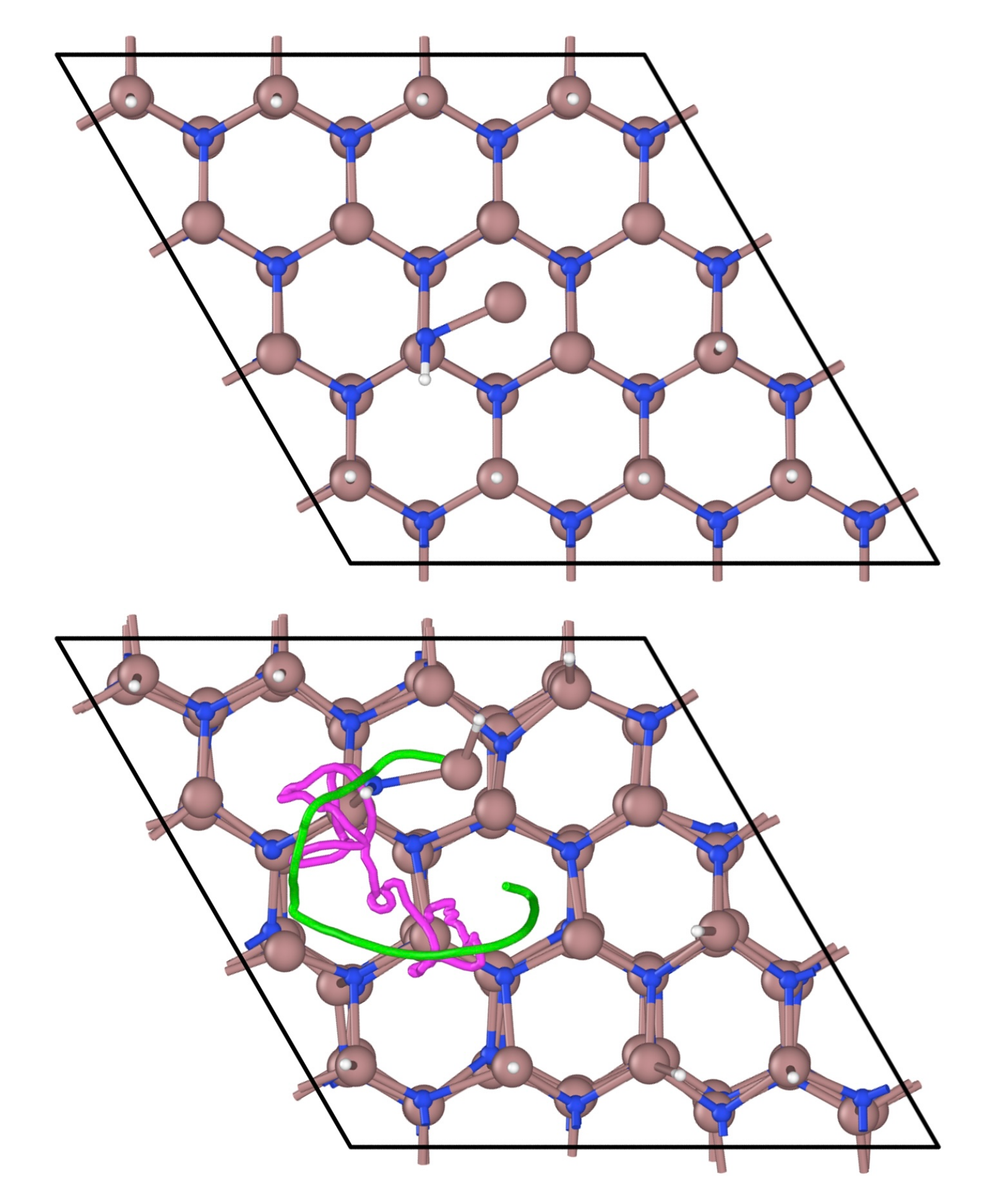}
\caption{Top views of the GaN(0001) slab model (Ga: brown, N: blue,
H: white); the black frame denotes the $4 \times 4$ supercell.
Top: initial configuration with the GaNH admolecule and nine H
adatoms.
Bottom: migration trajectories of Ga$_{\mathrm{admol}}$
(green) and the N atom of the GaNH unit (magenta) during the
1.5~ps FPMD run, superposed on the final configuration.}
\label{fig:initial-traj}
\end{figure}

The GaN(0001) surface was modeled by a periodic slab consisting of
five GaN bilayers with a $4 \times 4$ lateral cell and a vacuum region
of $\approx$16~\AA.
The bottom N layer was terminated by hydrogen atoms at a coverage of
3/4 monolayer, which satisfies the electron-counting rule, and the
bottom N layer and the terminating hydrogens were kept fixed at their
bulk positions throughout the simulations.
On the top surface, nine H adatoms and a GaNH admolecule were placed.
This composition corresponds to the hydrogen coverage of the 3Ga--H
$(2 \times 2)$ reconstruction---the stable surface under H$_2$ carrier
gas \cite{Kangawa2001,Kusaba2017JJAP}---maintained over three quarters
of the supercell, with the GaNH admolecule accounting for the area of
the remaining $(2 \times 2)$ cell.
The Ga atom belonging to the GaNH admolecule is hereafter denoted
Ga$_{\mathrm{admol}}$, to be distinguished from Ga adatoms
(Ga$_{\mathrm{ad}}$) and the Ga atoms of the crystalline surface.
The initial configuration is shown in the top panel of
Fig.~\ref{fig:initial-traj}; the geometry of the GaNH admolecule was
obtained by structural optimization prior to the MD runs.
This configuration models the growing surface immediately after
NH$_3$ decomposition under H$_2$-rich growth conditions, in which the
GaNH growth unit coexists with surface hydrogen.

MD runs were performed in the NVT ensemble with a Nos\'e--Hoover chain
thermostat at 1273~K, a typical MOVPE growth temperature, with a time
step of 0.1~fs, for a total simulation time of 1.5~ps (15{,}000 steps),
starting from the optimized geometry with initial velocities assigned
according to the target temperature.

\subsection*{Machine-learning interatomic potential}

For long-time simulations we employed UMA \cite{UMA2025}, a universal
machine-learning interatomic potential pretrained on $\sim$500 million
DFT calculations covering materials, molecular, and catalysis domains.
The UMA-m~1.1 model with the OC20 task head was used as implemented in
the FAIR-Chem package (fairchem-core 2.19.0), interfaced with LAMMPS
\cite{LAMMPS} (version 2~Aug~2023, update~3) through its
\texttt{fix external} mechanism (fairchem-lammps 0.4.0).
Prior to the production run, the initial structure was relaxed with UMA
using the FIRE algorithm \cite{FIRE} (force convergence 0.05~eV/\AA) in
ASE \cite{ASE}, and the relaxed structure was taken as the $t = 0$
configuration.

The accuracy of UMA for the present system was examined by single-point
recalculation of snapshots extracted from the FPMD trajectories:
751 snapshots sampled every 2~fs, from the initial
structure to the end of the 1.5~ps trajectory, were evaluated, and the
relative potential energies (per atom, referenced to the initial
structure) were compared with the CASTEP values.
In the analysis of the parity data, each snapshot was assigned to one of
the two branches according to the residual
$\Delta E_{\mathrm{MLIP}} - \Delta E_{\mathrm{DFT}}$ smoothed with a
7-point (14~fs) centered rolling mean, using a threshold of
6~meV/atom; the structural character of each branch was subsequently
identified from the instantaneous distance between
Ga$_{\mathrm{admol}}$ and the nearest top-layer Ga atom.

UMA-based MD simulations were then performed for the same slab model at
1273~K with a time step of 0.1~fs for a total of 150~ps
(1{,}500{,}000 steps), i.e., one hundred times longer than the FPMD
run.
To avoid thermostatting the adspecies dynamics directly, a mixed
integration scheme was adopted: the adspecies and the topmost GaN
bilayer were integrated in the NVE ensemble, whereas the second to
fourth bilayers and the Ga atoms of the bottom bilayer were coupled to
a Nos\'e--Hoover chain thermostat at 1273~K (chain length 5, damping
time 10~fs); the bottom N layer and the terminating hydrogens were
fixed, as in the FPMD runs.
In the analysis of the UMA trajectory, configurations were sampled
every 10~fs; the bonding state of Ga$_{\mathrm{admol}}$ was classified
by the structural criterion identified in the parity analysis
(Ga--Ga distance threshold of 3.3~\AA), with the dissociated periods
excluded.
Dissociation and re-formation events of the GaNH unit were identified
by visual inspection of the trajectory and delineated using the time
evolution of the distance between Ga$_{\mathrm{admol}}$ and the N atom
of the unit.
Atomic configurations and trajectories
(Figs.~\ref{fig:initial-traj}, \ref{fig:fpmd-lift},
\ref{fig:fpmd-habst}, and \ref{fig:umamd-traj}) were visualized using
OVITO \cite{OVITO}.

% ==========================================================================
\section*{Results and discussion}
% ==========================================================================

\subsection*{Dynamic behavior of the GaNH unit and H adatoms observed by FPMD}

\begin{figure}[!tb]
\centering
\includegraphics[width=\linewidth]{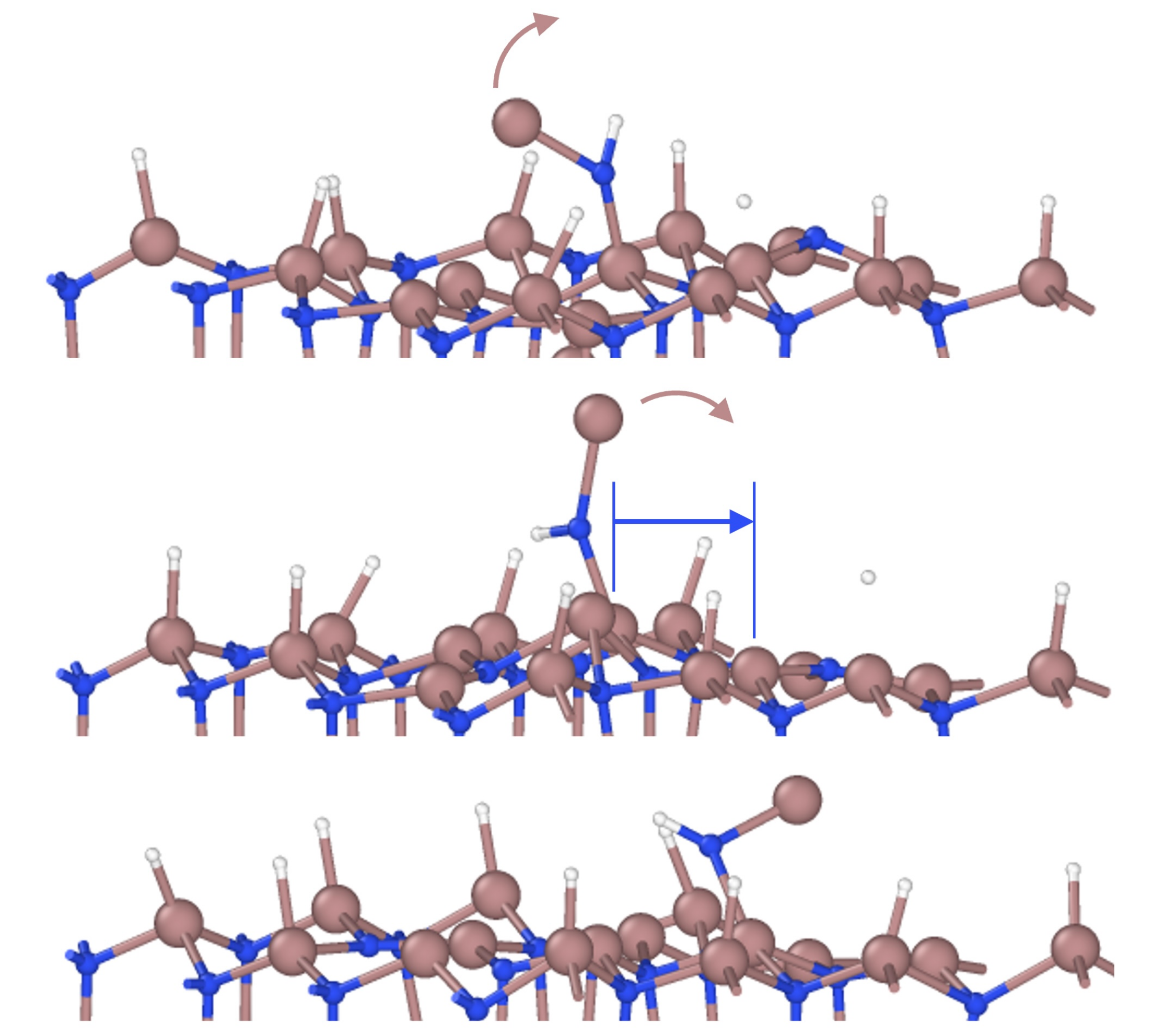}
\caption{Side-view snapshots from the FPMD trajectory showing the
lifting-assisted migration of the GaNH admolecule on GaN(0001)
(Ga: brown, N: blue, H: white).
The N atom of the GaNH unit lifts Ga$_{\mathrm{admol}}$ off the topmost
surface layer (top panel) and the unit migrates while holding it in the
elevated configuration (middle panel), after which the unit
settles at a new site (bottom panel).
Snapshots at $t = 0.2$, 0.4, and 0.8~ps (top to bottom) from the FPMD
run at 1273~K.}
\label{fig:fpmd-lift}
\end{figure}

The FPMD trajectories reveal a characteristic diffusion mode of the GaNH
admolecule.
As shown in Fig.~\ref{fig:fpmd-lift}, the GaNH unit does not simply hop
between adsorption sites as a rigid admolecule; instead, its N atom
\emph{lifts} Ga$_{\mathrm{admol}}$ off the topmost surface layer,
breaking the weak Ga--Ga contacts between Ga$_{\mathrm{admol}}$ and the
surface Ga atoms, and the unit migrates while holding the Ga in this
elevated configuration.
In the lifted configuration, Ga$_{\mathrm{admol}}$ resides typically
$\approx$3.3~\AA\ above the topmost Ga layer (up to $\approx$4.5~\AA),
compared with $\approx$2.5~\AA\ when it leans toward the surface; the
two configurations alternated about ten times during the 1.5-ps
trajectory, with residence times of up to $\approx$0.35~ps in either
configuration.
The resulting in-plane migration paths of Ga$_{\mathrm{admol}}$ and of
the N atom of the unit over the entire 1.5-ps trajectory are shown in
the bottom panel of Fig.~\ref{fig:initial-traj}.
This lifting-assisted migration was not---and arguably could not
be---identified by static DFT pathway searches, which presuppose a rigid
surface template; it is qualitatively consistent with the picture that
the Ga-rich growth front is dynamically disordered at growth temperature
\cite{Bui2020JJAP}.

\begin{figure}[!tb]
\centering
\includegraphics[width=6.3cm]{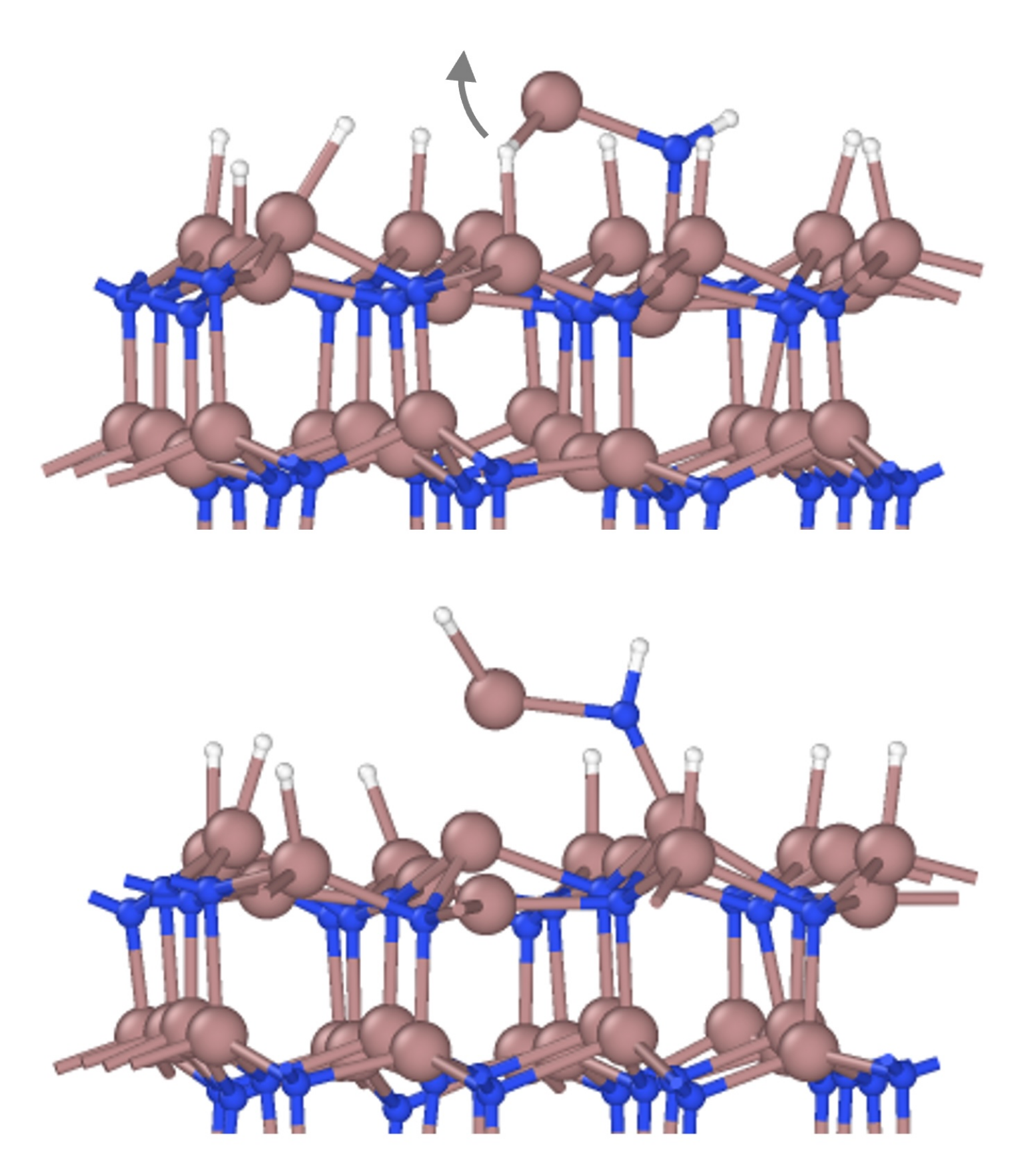}
\caption{Side-view snapshots from the FPMD trajectory showing hydrogen
abstraction by the lifted Ga$_{\mathrm{admol}}$ (Ga: brown, N: blue,
H: white).
The lifted Ga$_{\mathrm{admol}}$ captures a neighboring H adatom
(top panel) and abstracts it from the surface (bottom panel).
Snapshots at $t \approx 1.41$ and 1.50~ps (top and bottom) from the
FPMD run at 1273~K.}
\label{fig:fpmd-habst}
\end{figure}

Moreover, the lifted Ga$_{\mathrm{admol}}$ was observed to interact
with the coexisting H adatoms: as shown in Fig.~\ref{fig:fpmd-habst},
Ga$_{\mathrm{admol}}$ captures one of them and abstracts it from the
surface near the end of the trajectory ($t \approx 1.41$--$1.50$~ps),
yielding an H--Ga--NH admolecular complex detached from the H
adsorption site.
This provides a direct dynamical example of hydrogen removal from the
growing surface mediated by mobile Ga, complementing the
finite-temperature picture of hydrogen as a passivating but kinetically
active species \cite{Boero2022APSUSC}.

\subsection*{Validation of the UMA potential against first-principles data}

\begin{figure}[!tb]
\centering
\includegraphics[width=\linewidth]{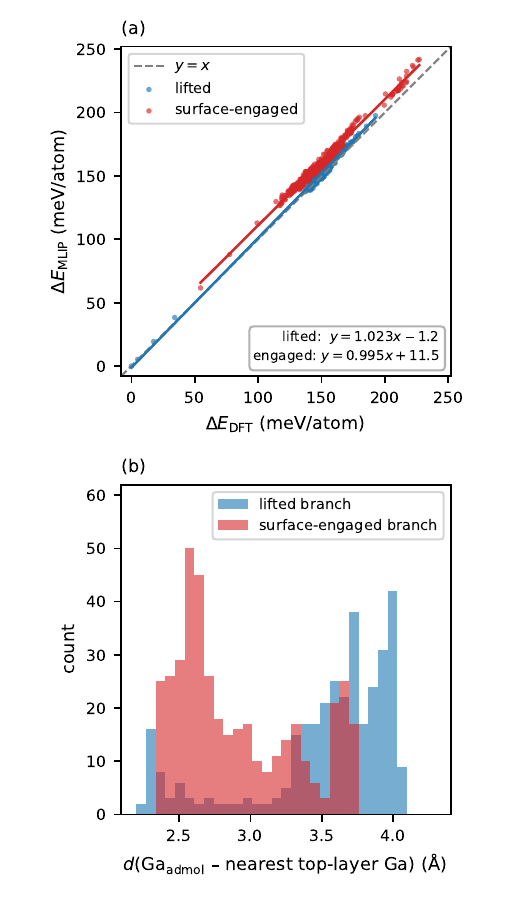}
\caption{Validation of the UMA potential against first-principles data,
resolved by the bonding state of Ga$_{\mathrm{admol}}$.
(a) Parity plot of the relative potential energies (in meV/atom,
referenced to the initial structure) of 751 snapshots sampled every 2~fs
along the 1.5~ps FPMD trajectory, computed by CASTEP
($\Delta E_{\mathrm{DFT}}$) and by single-point UMA calculations
($\Delta E_{\mathrm{MLIP}}$).
The data resolve into two branches: the \emph{lifted} branch (blue;
$y = 1.023x - 1.2$) and the \emph{surface-engaged} branch (red;
$y = 0.995x + 11.5$); the dashed line denotes $y = x$.
(b) Distributions of the distance between Ga$_{\mathrm{admol}}$ and the
nearest top-layer Ga atom for the two branches, showing that the lifted
branch is dominated by configurations without Ga--Ga contact
($d \gtrsim 3.3$~\AA) and the surface-engaged branch by configurations
with weak Ga--Ga bonds ($d \approx 2.5$--$3.0$~\AA).}
\label{fig:validation}
\end{figure}

Figure~\ref{fig:validation} compares the relative potential energies of
the 751 FPMD snapshots recalculated with UMA against the CASTEP
reference values.
The UMA energies accurately reproduce the first-principles relative
energetics along the trajectory: the parity plot yields an RMSE of
8.5~meV/atom with a correlation coefficient of 0.972 ($R^2 = 0.946$),
and the linear fit is essentially the identity line (slope 1.001) with
only a small positive offset of $\sim$7~meV/atom.
It is noteworthy that this accuracy is achieved \emph{without any
system-specific retraining}: the trajectory samples chemically demanding
configurations---weak metallic Ga--Ga bonds, under-coordinated adatoms,
N--H and Ga--H species, and thermally distorted lattices at growth
temperature---yet the pretrained universal potential describes their
relative energies at near-first-principles fidelity.
A closer inspection of the parity plot reveals that the residual
$\Delta E_{\mathrm{MLIP}} - \Delta E_{\mathrm{DFT}}$ is not uniform but
depends systematically on the bonding state of Ga$_{\mathrm{admol}}$:
the data points form two nearly parallel branches with almost identical
slopes.
A structural analysis of all 751 snapshots
[Fig.~\ref{fig:validation}(b)] shows that the branch closer to the
identity line (residual $\approx$ 1--3~meV/atom; fitted as
$y = 1.023x - 1.2$, in meV/atom) corresponds to the \emph{lifted}
configurations,
in which Ga$_{\mathrm{admol}}$ is held up by the N atom without any
Ga--Ga contact with the topmost surface layer, whereas the upper branch
(residual $\approx$ 10~meV/atom; $y = 0.995x + 11.5$) corresponds to
the \emph{surface-engaged} configurations, in which
Ga$_{\mathrm{admol}}$ leans toward the surface and forms \emph{weak
Ga--Ga bonds} with the top-layer Ga atoms (Ga--Ga distances of
$\sim$2.5--3.0~\AA).
The two branches are thus a direct fingerprint of the two bonding states
that alternate during the lifting-assisted migration described above.
In other words, the dominant part of the UMA error is a state-dependent
offset associated with the description of the weak metallic Ga--Ga
interaction---precisely the bonding motif that characterizes the
growing GaN surface \cite{Bui2018JPCC,Bui2021APSUSC}---while within each
bonding state the relative energy landscape follows the DFT reference
with a near-unity slope.
The magnitude of this offset ($\lesssim$10~meV/atom) sets the accuracy
scale for energy differences between the two bonding states, whereas
energy variations within a given state are reproduced with considerably
smaller errors.
This validation underpins the long-time simulations described below.

\subsection*{Long-time dynamics revealed by MLIP-based MD}

In the MLIP-based MD run, the GaNH unit exhibited the same
lifting-assisted migration mode as observed by FPMD, and did so
consistently throughout the 150-ps trajectory.
This agreement works in both directions: it provides a dynamical
consistency check on UMA that complements the energetic validation
above, and, conversely, it establishes that the lifting-assisted
migration is not a fortuitous episode of the short FPMD run but the
persistent migration mode of the GaNH unit at growth temperature.
The populations of the lifted and surface-engaged configurations
sampled in the 150-ps MLIP-based MD trajectory (56\% and 44\%, using the
structural criterion of Fig.~\ref{fig:validation}(b)) are similar to
those in the FPMD trajectory (50\% and 50\%), suggesting that the
state-dependent offset identified above does not grossly bias the
configurational sampling---although the short FPMD run provides only
limited reference statistics.

\begin{figure}[!tb]
\centering
\includegraphics[width=\linewidth]{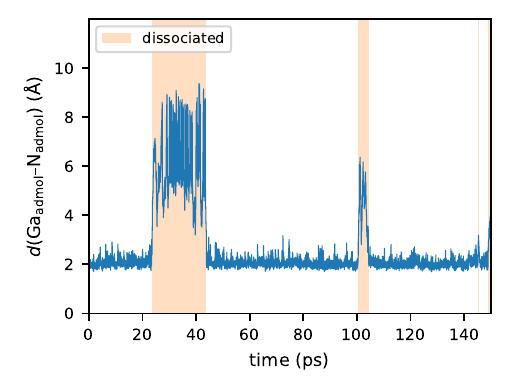}
\caption{Reversible dissociation and re-formation of the GaNH unit
observed in the 150-ps MLIP-based MD run.
The distance between Ga$_{\mathrm{admol}}$ and the N atom of the unit
is shown as a function of time (sampled every 10~fs, minimum-image
convention).
Shaded intervals denote the dissociated periods, in which the unit is
split into a Ga adatom and an NH admolecule migrating independently
over the surface (dissociation at 23.4, 100.5, 145.2, and 149.0~ps;
re-formation at 43.9, 104.6, and 145.7~ps).}
\label{fig:umamd-dissoc}
\end{figure}

\begin{figure}[!tb]
\centering
\includegraphics[width=\linewidth]{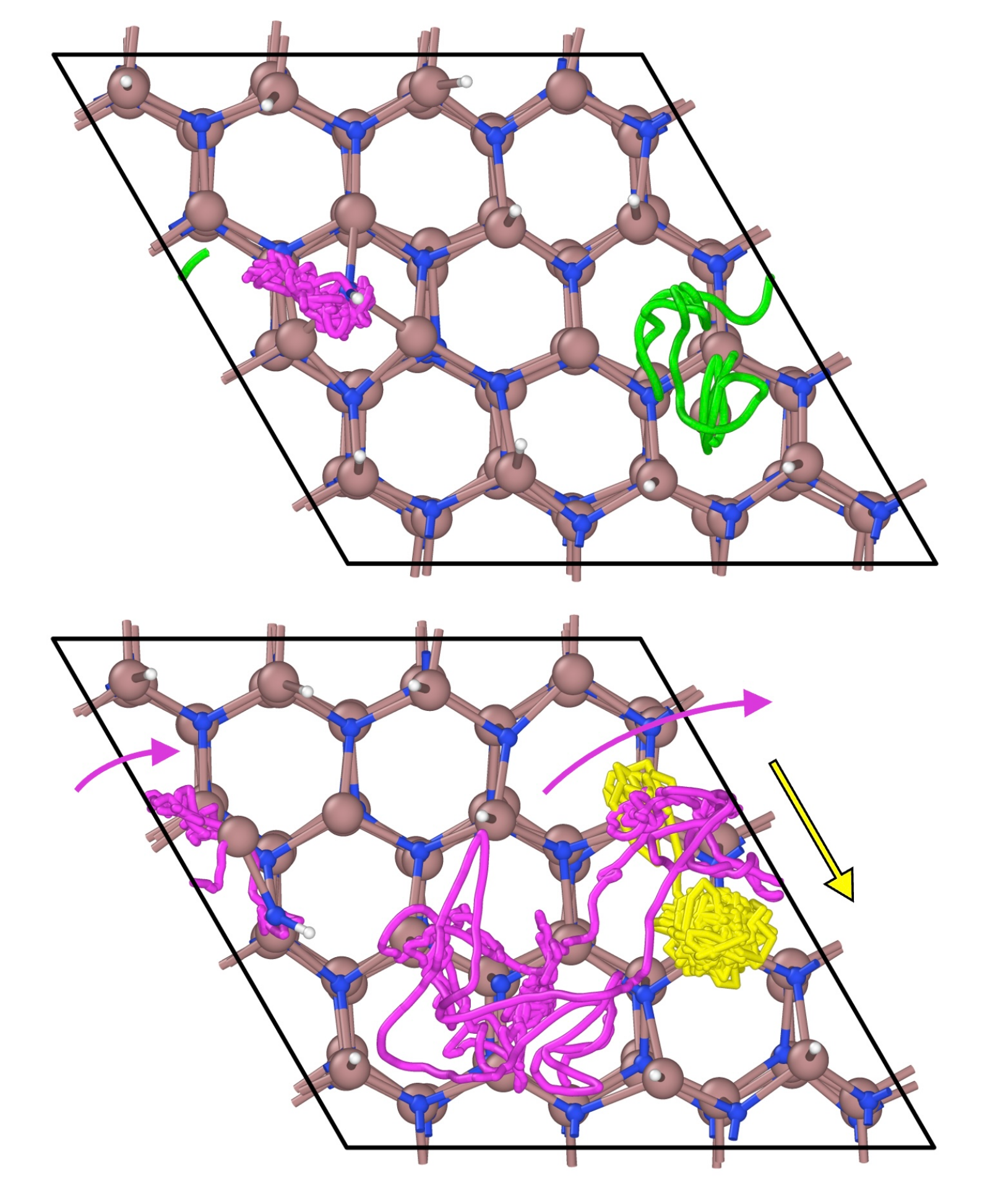}
\caption{Top views of migration trajectories from the MLIP-based MD run
(Ga: brown, N: blue, H: white); the black frame denotes the
$4 \times 4$ supercell.
Top: dissociated period following the first dissociation event
(23.4--30~ps); the Ga adatom (green) and the NH admolecule (magenta;
trajectory of its N atom) migrate independently over the surface.
Bottom: hydrogen-gated migration of the GaNH unit (105--115~ps); the
hop of an H adatom (yellow, arrow) opens a path through which the GaNH
unit subsequently migrates (magenta, arrows).}
\label{fig:umamd-traj}
\end{figure}

The extended time window accessible with the MLIP (150~ps) reveals
elementary processes that FPMD could not capture.
Already the first site-to-site hop of an H adatom occurred at 3.4~ps,
just beyond the 1.5-ps span of the FPMD run, and such hops recurred
throughout the trajectory; their role in gating the migration of the
growth unit is discussed below.
Most importantly, the GaNH unit is found to be a \emph{reversibly
dissociating} species.
As summarized in Fig.~\ref{fig:umamd-dissoc}, four dissociation events
occurred during the 150-ps run (at 23.4, 100.5, 145.2, and 149.0~ps):
the unit dissociates into a Ga adatom and an NH admolecule, which
migrate independently over the surface
(Fig.~\ref{fig:umamd-traj}, top) for periods ranging from
$\sim$0.4 to $\sim$20~ps and eventually re-encounter each other to
re-form the GaNH unit; the final dissociation had not yet recombined
when the run ended.
The dynamical equilibrium
GaNH $\rightleftharpoons$ Ga$_{\mathrm{ad}}$ + NH$_{\mathrm{ad}}$
implies that mass transport of the growth unit on the surface proceeds
not as a rigid admolecule but through repeated dissociation and
re-association.
This picture dynamically generalizes the concerted
(``hand-in-hand'') NH--Ga migration proposed from static DFT pathway
analysis \cite{Bui2023APSUSC}: at growth temperature the partnership
between NH and Ga is intermittent rather than permanent, with direct
implications for the length scale of adspecies migration prior to
incorporation at steps \cite{Bui2021APSUSC,Bui2023APSUSC}.

The extended run also revealed a kinetic interplay between surface
hydrogen and the transport of the growth unit.
Because the N atom of the unit binds to the topmost-layer Ga atoms,
the GaNH unit migrates through H-free surface sites, and hops of H
adatoms between adsorption sites---observed repeatedly during the
run---open and close its migration paths; in one instance, an H hop at
$\approx$105~ps opened a path through which the GaNH unit subsequently
migrated ($\approx$110--113~ps; Fig.~\ref{fig:umamd-traj}, bottom).
This provides a dynamical counterpart to the blocking of NH migration
by hydrogenated surface regions inferred from static pathway
calculations \cite{Bui2023APSUSC}.

We note that the hydrogen abstraction by the lifted
Ga$_{\mathrm{admol}}$ observed in the FPMD run did not recur within
the 150-ps MLIP-based MD trajectory, whereas single H adatom hops did
occur
repeatedly.
This suggests that the abstraction is a rare event contingent on a
specific encounter geometry between the lifted Ga$_{\mathrm{admol}}$
and a neighboring H adatom, which the FPMD trajectory happened to
sample; assessing its frequency---and hence its contribution to
hydrogen removal during growth---will require still longer simulations
or ensembles of trajectories.

% ==========================================================================
\section*{Conclusions}
% ==========================================================================

We have investigated the dynamic elementary processes of growth-related
adspecies on the GaN(0001) surface under MOVPE-relevant
conditions by combining first-principles molecular dynamics (CASTEP)
with the universal machine-learning interatomic potential UMA.
The main conclusions are as follows.
(i) FPMD simulations of a GaNH admolecule coexisting with H adatoms
revealed a lifting-assisted migration mode in which the N atom of the
growth unit lifts Ga$_{\mathrm{admol}}$ off the surface layer, and
showed that the lifted Ga$_{\mathrm{admol}}$ abstracts an H adatom
from the surface; neither event is accessible to static DFT approaches.
(ii) Single-point UMA calculations on FPMD snapshots reproduced the
first-principles relative energies along the trajectory
(RMSE $\approx$ 8.5~meV/atom) without any system-specific retraining,
validating the universal potential for the GaN growth surface.
(iii) Long-time MLIP-based MD (150~ps) revealed dynamic behavior that
never occurred within the FPMD time window: H adatoms hop between
adsorption sites---opening and closing the migration paths of the
growth unit---and the GaNH unit reversibly dissociates into
independently migrating Ga and NH adspecies that later re-encounter
and re-form the unit.
Taken together, these results recast the picture of GaN growth built up
by previous static DFT studies---the GaNH complex as the elementary
growth unit, transported via the concerted (``hand-in-hand'')
migration of NH with a Ga adatom, and surface hydrogen as a blocker of
migration---in explicitly dynamical terms: at growth temperature, the
partnership between Ga and NH is intermittent rather than permanent,
and surface hydrogen gates, rather than statically blocks, the
transport of the growth unit.

To our knowledge, this work is the first application of a
machine-learning interatomic potential to the molecular dynamics of
GaN MOVPE growth.
The demonstrated combination---first-principles benchmarking of a
universal potential followed by long-time MLIP dynamics---provides a
practical route to bridge the time-scale gap between ab initio accuracy
and growth-relevant kinetics, and is expected to substantially advance
the atomistic understanding of GaN MOVPE growth mechanisms, including
precursor kinetics, hydrogen dynamics, and ultimately step-flow growth
and defect formation.

% ==========================================================================
% Back matter
% ==========================================================================
\section*{Acknowledgements}
The authors thank Tetsuji Kuboyama for his support regarding the use of UMA.
This work was supported by JSPS KAKENHI (grant numbers JP23K28151, JP24K17619, and JP24H00432) and JST BOOST (grant number JPMJBY24C3).
The computation was carried out using the computer resource offered under the category of General Projects by Research Institute for Information Technology, Kyushu University.

\section*{Author contributions statement}
Y.T. performed the UMA-based simulations; A.K. conceived and
supervised the study and wrote the manuscript; J.I. and S.M. performed
the FPMD simulations; Y.K. supervised the project.
All authors reviewed the manuscript.

\end{document}